\documentclass[10pt]{article}
\usepackage{amsmath}
\usepackage{amssymb}
\usepackage{graphicx}

\def\gji{Geophys. J. Int.}

\def\jgr{J. Geophys. Res.}

\begin{document}

\setcounter{figure}{0}
\setcounter{table}{0}
\setcounter{footnote}{0}
\setcounter{equation}{0}

\vspace*{0.5cm}

\noindent {\Large \strut ON DETECTION OF THE FREE INNER CORE NUTATION FROM VLBI DATA}
\vspace*{0.7cm}

\noindent\hspace*{1.5cm} Z. MALKIN\\
\noindent\hspace*{1.5cm} Pulkovo Observatory, St. Petersburg, Russia\\
\noindent\hspace*{1.5cm} St. Petersburg State University, St. Petersburg, Russia\\
\noindent\hspace*{1.5cm} e-mail: malkin@gao.spb.ru\\

\vspace*{0.5cm}

\noindent {\large ABSTRACT.}
Several attempts to discover the FICN signal in VLBI nutation series made during last years failed.
In this paper, we present some results of our further steps in this direction, unfortunately not successful either.
We investigated several VLBI CPO series by means of spectral and wavelet analysis.
It has been shown that there are several periodic signals with close amplitude around the expected FICN period without any prevailing one that
can be reliably associated with the FICN.
The most interesting for further analysis is a relatively stable oscillation with period of 800 days, which is, however, beyond the intervals predicted in other
studies cited above.
It seems to be necessary to improve  theoretical estimates of the FICN period to make its search in the observational data more promising.

\vspace*{1cm}

\noindent {\large 1. INTRODUCTION}

\smallskip

Free inner core nutation (FICN,) is one of the four free rotational modes of the Earth considered in the theory of the Earth's rotation.
Detecting of this signal in the observational data is a very important scientific task allowing us to substantially improve our knowledge
about the Earth's interior and dynamics.

According to Mathews et al. (2002) the FICN period is between 930 and 1140 days.
Koot et al. (2010) estimated the FICN period from $904 \pm 29$ to $945 \pm 30$, i.e. between 875 and 975
days (1$\sigma$ interval).
Because of a small expected amplitude of the  FICN signal its detection can be successful only from the most accurate nutation series obtained
from the VLBI observations.
It may be also possible that the FICN oscillation has the amplitude and phase variations like free core nutation (FCN).

Several attempts made during last years to find the FICN component in these series failed, see, e.g., Lambert et al. (2012) and papers cited therein.
Moreover, the results depend on the celestial pole offset (CPO) series ($dX, dY$) used.
In this work, we performed a new analysis of all available CPO series to investigate possible geophysical signals in expected FICN frequency band.

\vspace*{0.7cm}

\noindent {\large 2. DATA ANALYSIS}

\smallskip

We analyzed several CPO series obtained in IVS analysis centers by means of spectral and wavelet analysis.
These series include the combined IVS series, as well as individual CPO series obtained in IVS Analysis Centers:
AUS (Australia), BKG (Germany), CGS (Italy), GSF (USA), IAA (Russia), OPA (France), USN (USA). 

As it was shown in many studies, the main components of CPO include long-term trend caused by the errors in modelling precession
and low-frequency nutation terms, and free core nutation (FCN).
The FCN signal was removed from the CPO time series prior to further analysis.

\begin{figure}[ht!]
\centering
\includegraphics[clip,width=0.9\textwidth]{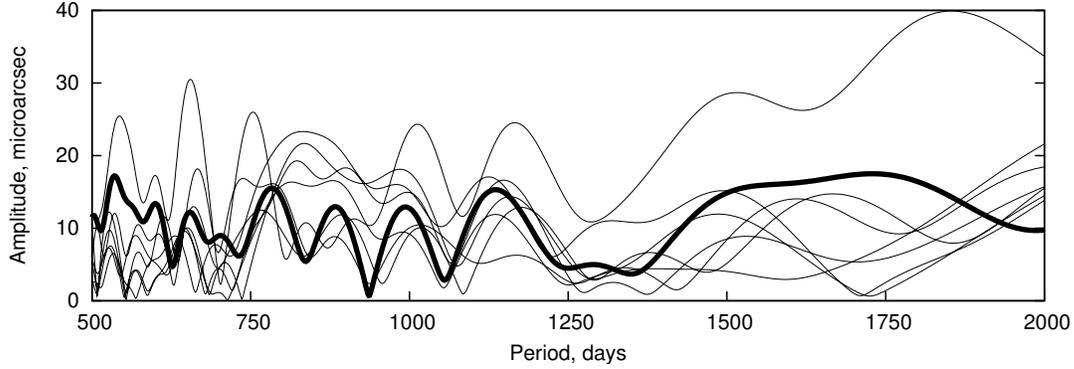}
\caption{Periodogram of the individual (think line) and IVS combined (thick line) CPO series after removing FCN.}
\label{fig:spectra}
\end{figure}

Firstly, the spectral analysis was applied to all CPO series in complex form $dX + i dY$.
Result of these computations is presented in Fig.~\ref{fig:spectra}.
One can see from these spectra that the CPO variations in the FICN frequency band show several unstable harmonics of similar amplitude,
and no prevailing signal is revealed that can be reliably associated with the FICN.
Correlation between the spectra of different series is not very good.
Properly speaking, agreement between the CPO series is, in fact, rather poor, which does not allow one to reliably obtain the signal pattern. 
Some correlation between series can be explained by both presence of real physical signal and using the same observational data and similar analysis options.

\begin{figure}[ht!]
\centering
\includegraphics[clip,width=0.9\textwidth]{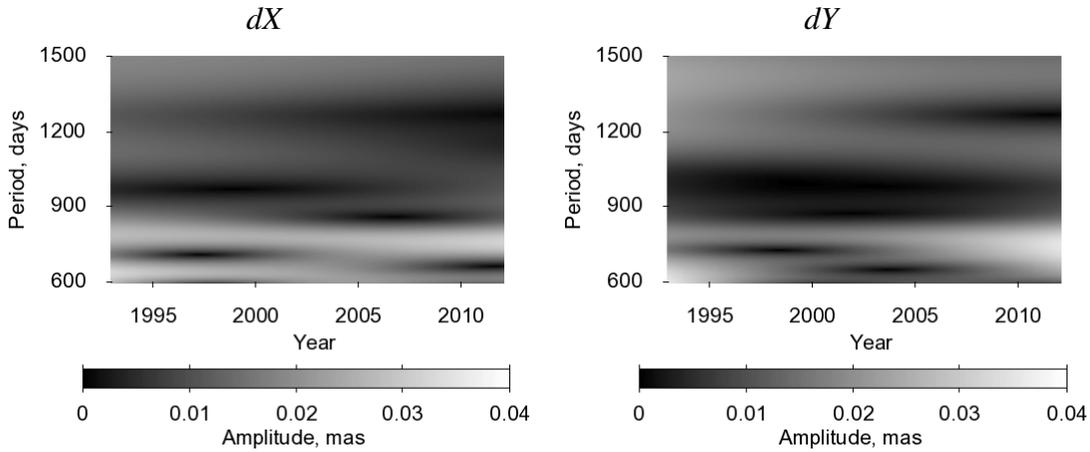}
\caption{Wavelet scalograms for the IVS combined CPO series after removing FCN.}
\label{fig:wavelet}
\end{figure}

Figure~\ref{fig:wavelet} shows the result of wavelet analysis applied to the IVS combined CPO series after removing trend and FCN.
One can see a complicated structure of the CPO variations without a clearly detected signal around the expected FICN period,
except the signal with period about 800 days.

\clearpage

\vspace*{0.4cm}

\noindent {\large 4. CONCLUSION}

\smallskip
The results of this study presented above show that the CPO variations in the FICN frequency band have a complicated structure, as was already
shown by Lambert et al. (2012).
Several unstable harmonics of close amplitude are present in the spectra, and no prevailing signal can be reliably associated with the FICN.
Wavelet analysis revealed relatively stable oscillation with period of 800 days, which is, however, beyond the intervals found in other studies cited above.
Unfortunately, the theoretical prediction of the FICN period is not sufficiently accurate to unambiguously connect one of the oscillations with FICN.
Hence, it seems to be necessary to improve  the theoretical estimates of the FICN period to make its search in the observational data more promising.

\bigskip
\noindent {\it Acknowledgements.} The author is grateful to the organizers of the conference for the travel support.

\vspace*{0.7cm}

\noindent {\large 5. REFERENCES}

{

\leftskip=5mm
\parindent=-5mm

\smallskip

Koot, L., Dumberry, M., Rivoldini, A., de Viron, O., Dehant, V., 2010, ``Constraints on the coupling at the core-mantle and inner core
boundaries inferred from nutation observations'', \gji, 182, pp.~1279--1294.

Mathews, P.M., Herring, T.A., Buffett, B.A., 2002, ``Modeling of nutation and precession: New nutation series for nonrigid Earth
and insights into the Earth's interior'', \jgr, 107(B4), 2068.

Lambert, S., Rosat, S., Cui, X., Rogister, Y., Bizouard, C., 2012, ``A Search for the Free Inner Core Nutation in VLBI Data'', 
IVS 2012 General Meeting Proc., pp.370--374.

}

\end{document}